\documentclass[12pt]{iopart}
\usepackage{graphicx}
\usepackage{xcolor}
\usepackage{siunitx}

\usepackage{epstopdf}
\begin{document}

\title[Author guidelines for IOP Publishing journals in  \LaTeXe]{Teaching labs for blind students: equipment to measure standing waves on a string}

\author{A. Lisboa  $^{1}$*, F. J. Pe\~na$^{1}$, O. Negrete $^{1,2}$ and C.O. Dib$^{1,3}$}

\address{%
	$^{1}$ \quad Departamento de F\'isica, Universidad T\'ecnica Federico Santa Mar\'ia, Casilla 110-V 2390123 Valpara\'iso, Chile. \\
	$^{2}$ \quad Centro para el Desarrollo de la Nanociencia y la Nanotecnolog\'ia, 8320000 Santiago, Chile. \\
	$^{3}$ \quad Centro-Cient\'ifico-Tecnol\'ogico de Valpara\'iso, CCTVal, Valpara\'iso, Chile.}

\ead{alfredo.navarro@usm.cl, francisco.penar@usm.cl, oscar.negrete@usm.cl, claudio.dib@usm.cl}
\vspace{10pt}
\begin{indented}
\item[] May 2021
\end{indented}

\begin{abstract}

We designed a Physics Teaching Lab experience for blind students to measure the wavelength of standing waves on a string. Our adaptation consisted of modifying the determination of the wavelength of the standing wave, which is usually done by visual inspection of the nodes and antinodes, using the sound volume generated by a guitar pickup at different points along the string. This allows one of the blind students at our University to participate simultaneously as their classmates in the laboratory session corresponding to the wave unit of a standard engineering course.

\end{abstract}

\section{Introduction}

Seismic tremors, sounds, vibrations moving along a stretched string - these are all examples of mechanical waves. These can overlap, and a particular form of superposition corresponds to the generation of so-called standing waves. These are produced by the interference of two identical waves that travel in opposite directions. A simple way to generate them is to produce oscillations in a string that is fixed at its two ends. Waves generated at one end will travel along the string and are reflected at the other, generating a continuous superposition that gives rise to interference patterns characteristic of standing waves.

In our experimental setup, we are interested in determining the vibration's normal modes of a stretched string. A standing wave in a medium like a string, unlike a propagating wave, is a normal mode of oscillation of the medium. In a normal mode, all points in the medium oscillate together with the same frequency and, up to a sign, with the same phase. 

In general, the wavelengths $(\lambda_{n})$ of the different normal modes for a stretched string of length $L$ are given by \cite{Freedman,Serway}

\begin{equation}
    \label{normalmode}
    \lambda_{n}=\frac{2L}{n}, \quad n=1,2,3..
\end{equation}
where the index $n$ refers to the $n$-th normal mode of oscillation. The frequency of the corresponding normal mode is 
\begin{equation}
    \label{freq_n}
    f_n = \frac{v}{\lambda_n},
\end{equation}
where $v$ is the propagation speed of the wave in the string.

In our usual laboratory sessions, for a blind student it is impossible to do the actual measurements because the normal modes must be identified visually by noticing the formation of nodes (points where the string does not vibrate). 
Therefore, an adaptation of the device is necessary for a blind student to actively participate in the measurements \cite{Holt,DeBuvitz}. A natural alternative to visual readings is to include sounds related to the value of the physical variable 
 to be measured. 
In this context an open-source electronic prototyping platform (Arduino board) has been used in several recent experiments \cite{Marinho,Goncalves1,Atkin,KuvNob,Goncalves2,Toenders,Galeriu1,Galeriu2,Galeriu3} as it offers the possibility of using a variety of sensors and electronic components for the data acquisition and control of different experiments in physics.

In the literature, there is ample evidence of the great challenge that material adaptations constitute so that blind students can discuss, participate and feel included in conventional academic activities carried out by the rest of the students \cite{Millar,Dunkerton,Azevedo,Melaku,Fantin,Zou,Lewin,Scanlon}. However, making improvements so that blind students can participate in a physics laboratory is a requirement today.

In this work we present a successful adaptation for a blind student of an undergraduate-oriented physics laboratory experiment. The experiment is about determining the normal modes of oscillation of a stretched string. Our adaptation consisted in replacing the visual identification of nodes and antinodes in the string by a system where 
the node positions can be determined 
by an audible signal.

\section{Experimental setup}

\subsection{The standard setup}

\begin{figure}
    \centering
    \includegraphics[width=0.7\textwidth]{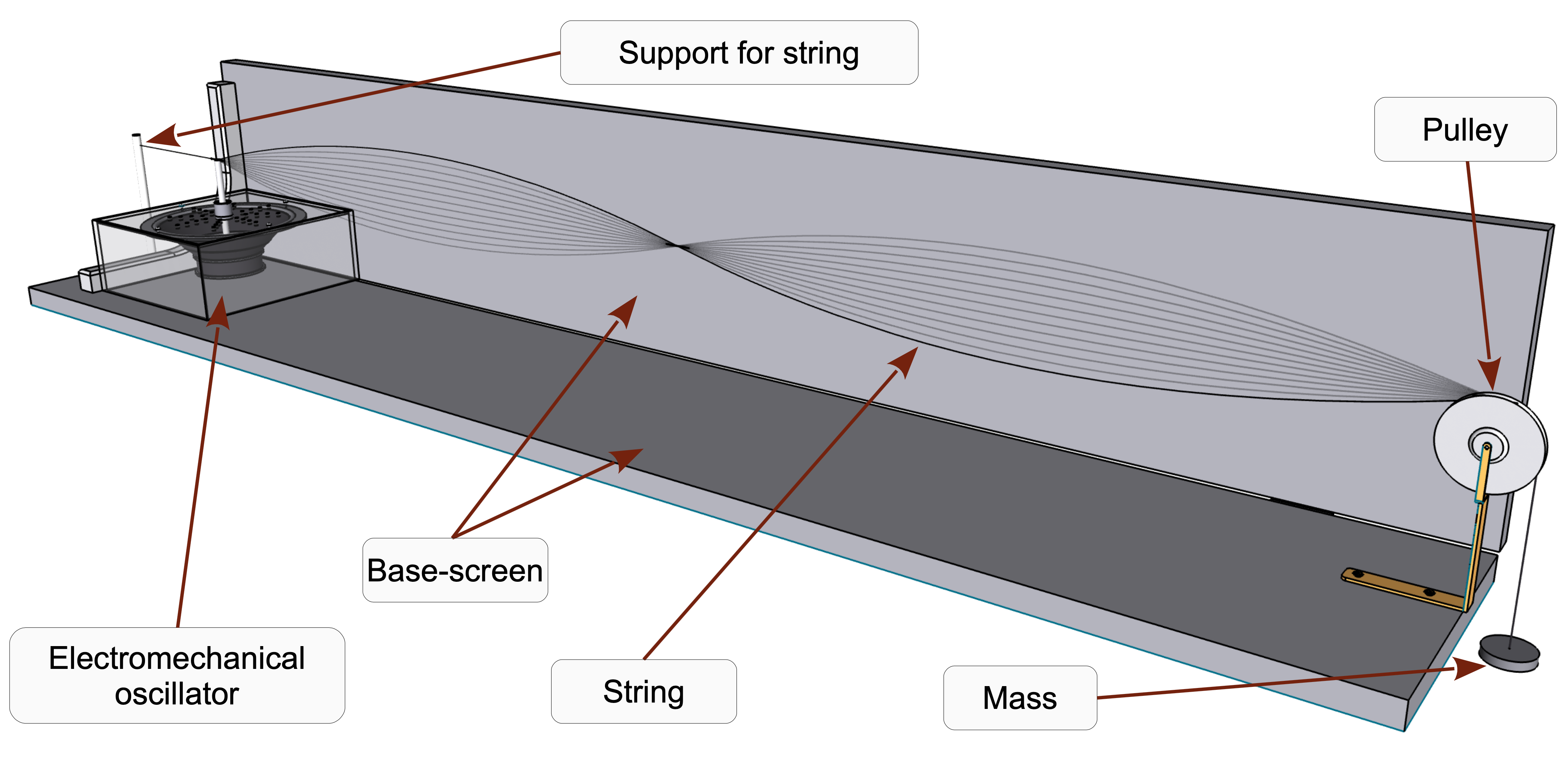}
    \caption{Schematic of the traditional experimental setup used in the laboratory of standing waves on a string.}
    \label{fig1}
\end{figure}

In this 
laboratory
{ session}, 
students must produce standing waves in a guitar string under tension and with fixed ends, by applying an excitation of adjustable frequency.


Fig.~\ref{fig1} shows our traditional equipment to study standing waves. The bottom base, 
about one meter long, is black to improve contrast with the metal string. The string is tied to a fixed support at one end and an electromechanical oscillator is connected to the string a few cm from the support. At the other end, the string hangs down over a pulley, where a mass is hanged to provide a measurable string tension.

The string tension is then simply given by the weight, i.e. $T=mg$.

The electromechanical oscillator that excites the string is a loudspeaker 
with a small rod that connects its central membrane to the string. The loudspeaker is connected to an amplifier, in turn connected to a computer.
The amplitude and frequency of the signal is
set at the computer 
with a signal generator software. 


The amplitude is set at a constant value
while the frequency is gradually increased at intervals of 0.1 Hz, starting from 0.25 Hz until the frequency of the string's 7th harmonic mode. 

The students are required to find the frequency values of the first five normal modes of the string. 
The mass, $m_{s}$, and total length of the string, $l$,  must be measured, and the linear mass density, $\mu=m_{s}/l$, calculated. Knowing the value of $\mu$ and the tension $T$ applied to the string, the value of the wave speed propagation in the string is determined with the expression 

\begin{equation}\label{vref}
v_{r}=\sqrt{\frac{T}{\mu}}.
\end{equation}
This speed we call  $v_{r}$, and must be compared with the speed obtained from the product of the frequency $f$ times the wavelength $\lambda$, which  we call $v_{ms}$:
\begin{equation}
    v_{ms}=\lambda f.
\end{equation}

\subsection{Our modified setup}

\begin{figure}
    \centering
    \includegraphics[width=0.7\textwidth]{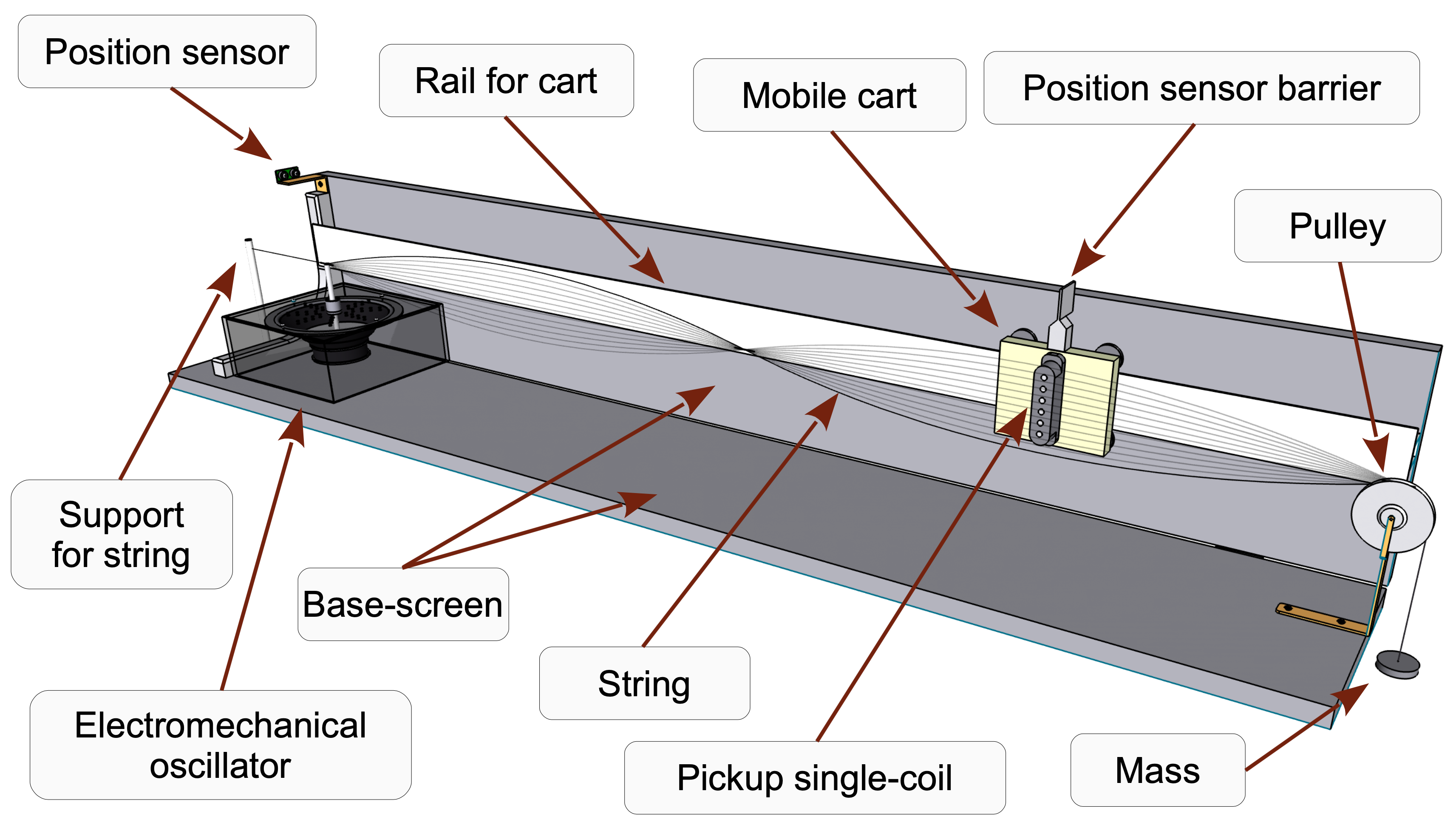}
    \caption{Design of the experimental system so that a blind student can measure physical quantities in a laboratory of standing waves on a string.}
    \label{fig2}
\end{figure}

In our standard setup described above, 
the modes are determined visually, simply by observing for what frequency the correct standing wave is formed, with its nodes and antinodes.

In order to enable a blind student to find the nodes and antinodes in the string, many alternative instruments were explored, such as light intensity sensors, photoresistors, and magnetic field sensors. However, the choice that showed the best results consisted in
replacing the nylon string with an metallic electric guitar string and a single-coil guitar pickup. The pickup is composed of a permanent magnet and a coil. When the string vibrates in front of the magnet, it disturbs the magnetic field generating induced currents in the coil. Guitar pickups use a permanent magnet for each of the guitar strings (6 in total). Now, in our setup, when the frequency corresponds to a normal mode of oscillation of the string, 
nodes and antinodes are formed, which can be found by sliding the pickup in front of them. When placed in front of a node, the signal in the pickup is minimal and in front of an antinode is maximal. The signal is transformed into sound through an amplifier, which is easily distinguishable to a blind student.

Then, the wavelength of the mode can be measured by the blind student, knowing that the distance between two consecutive nodes or consecutive antinodes corresponds to $\lambda/2$. 

The experiment also requires to read the frequency on the computer screen. While an adaptation for blind students can be done here too, we did not do that in order to promote collaborative work with other student. In this case the blind student's partner should not be visually impaired.


Figure \ref{fig2} shows the apparatus existing in our laboratory 
with the modifications to include the guitar pickup:
to the standard setup in the lab, we added a rail parallel to the string, along which a mobile cart can slide. The pickup is fixed at the center of the cart, facing the string. 


In order for the blind student to read the positions of the pickup and thus determine the distance between nodes and antinodes, a distance sensor with audio-reading is added.
This sensor is controlled by an Arduino platform.





\begin{figure}
    \centering
    \includegraphics[width=0.7\textwidth]{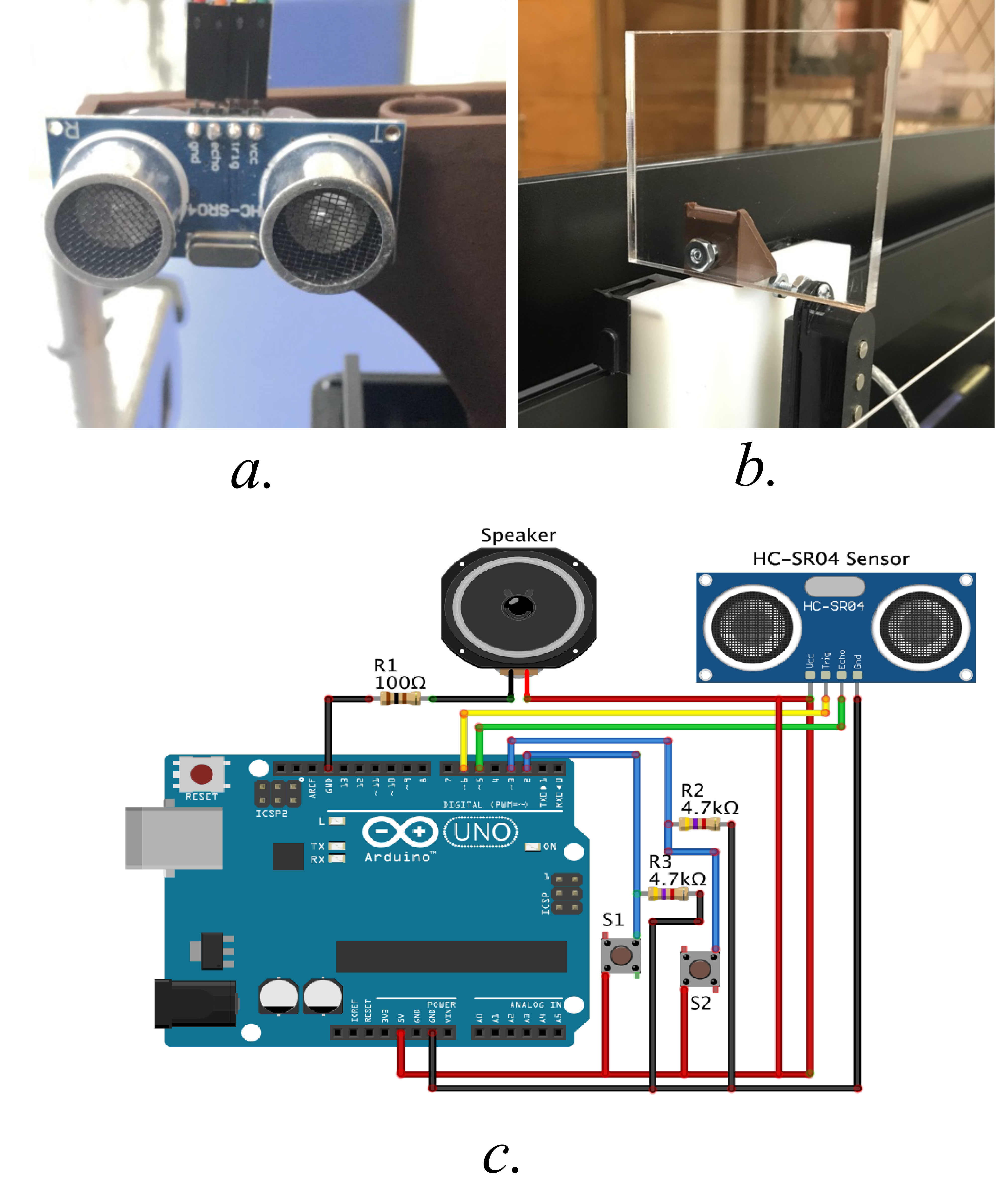}
    \caption{a. Position sensor compatible with Arduino platform. b. Barrier located on the mobile cart to determine the position of the capsule with respect to the origin of the coordinate system. c. Scheme of the electronic circuit for the position sensor.}
    \label{fig3}
\end{figure}

The position sensor is configured so that the zero position is at the point where oscillator touches the string. The sensor measures the distance from this point to an acrylic plate (a ``barrier'') mounted on the moving cart.

Figures \ref{fig3}(a) and \ref{fig3}(b) show the position sensor and the barrier placed on the cart, respectively. Two buttons, $S_1$ and $S_2$, were added to the circuit  [see figure \ref{fig3}(c)] to be managed independently by the blind student: $S_1$ restarts the measurement and $S_2$ delivers the positions through audio reading using \textit{Talkie library} software. The whole electronic circuit scheme developed for the Arduino platform can be seen in figure \ref{fig3}(c).

The mobile cart was manufactured in machinable plastic polyethylene, with size 98 mm long, 84 mm high and 15 mm thick. It has four wheels with grooved bearings to reduce friction with the rail and avoid sudden and inaccurate movements. The pickup is placed at the center of the cart, and the distance to the string is adjusted through two bolts. 

\begin{figure}
    \centering
    \includegraphics[width=0.7\textwidth]{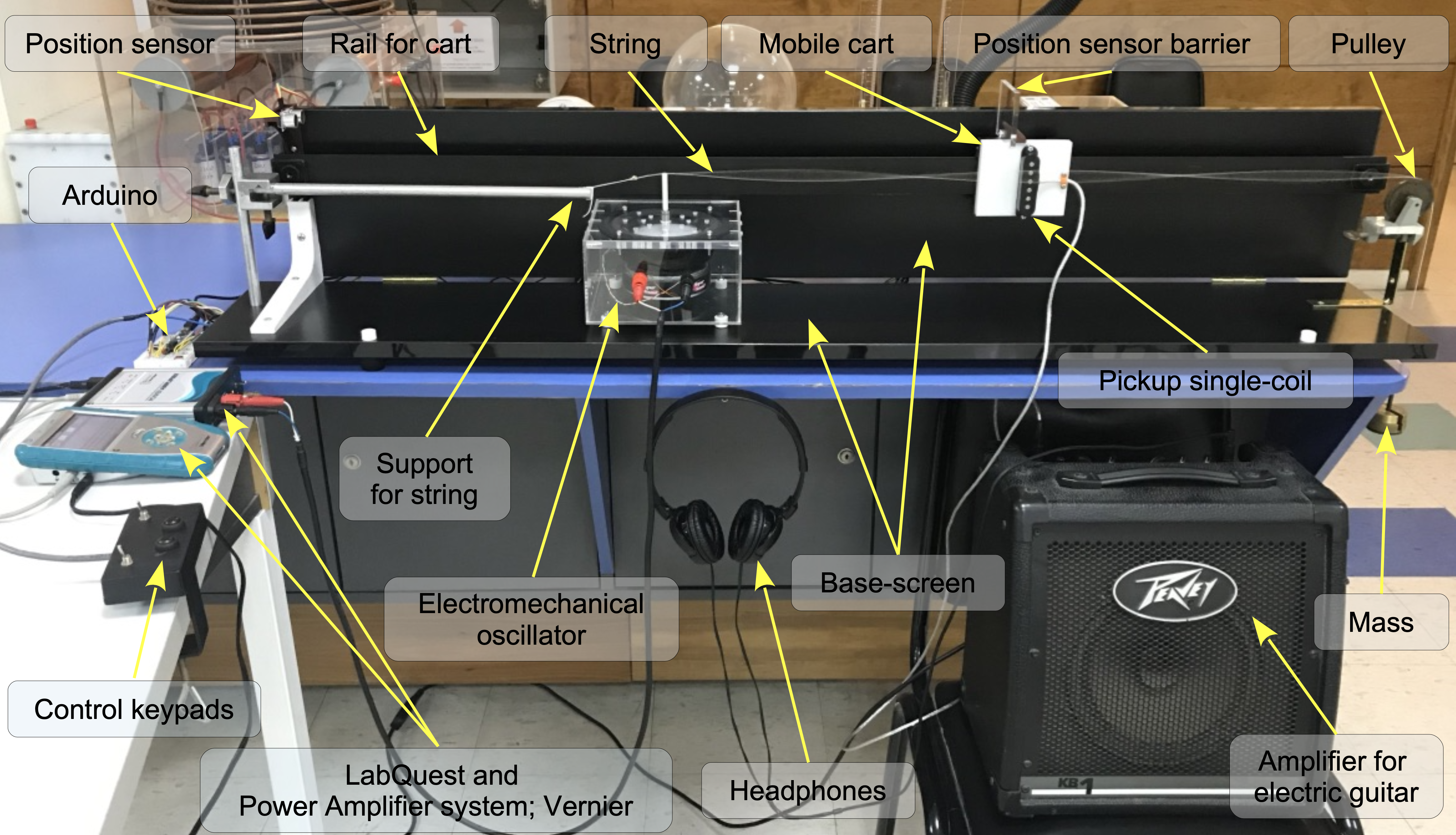}
    \caption{The full apparatus with the adaptations for blind students. The image shows the guitar string oscillating at its third harmonic mode. }
    \label{fig4}
\end{figure}

The mobile cart slides on an aluminum rail  1.25 m long, 43 mm high and 3 mm wide. It is 28 mm above the bottom and fastened to the side screen by two table presses, allowing the height to be adjusted and aligned along the string. The rail was leveled so that the string faces the third magnet of the pickup. To improve the identification of the nodes or antinodes, a guitar amplifier that feeds a headphone set, so that the blind student can better concentrate on the work.

\begin{figure}
    \centering
    \includegraphics[width=0.9\textwidth]{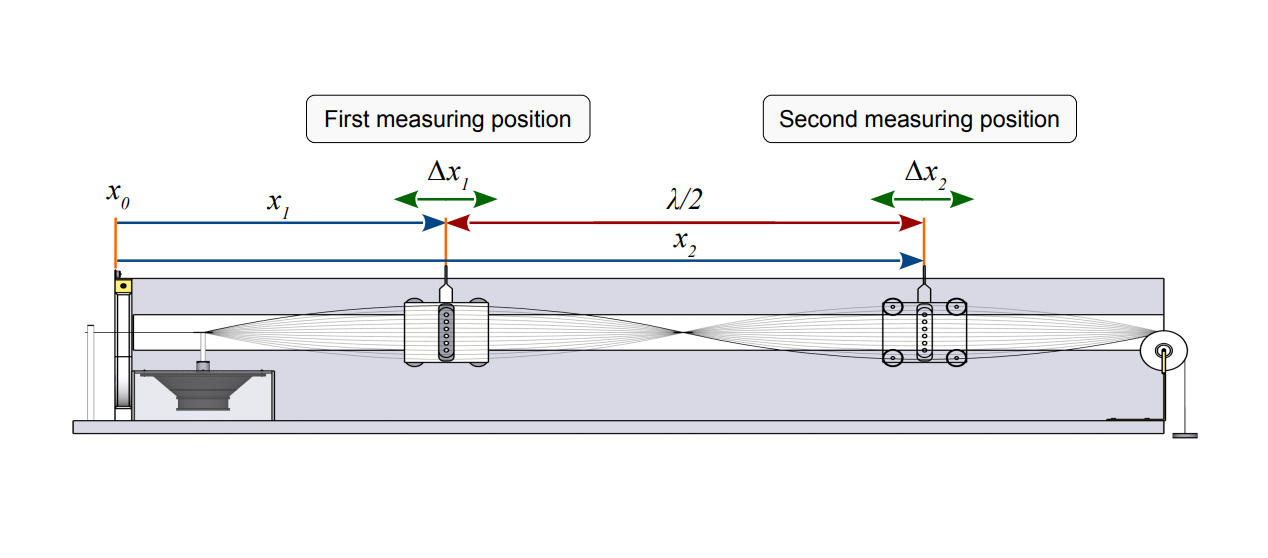}
    \caption{Measurement scheme of the modified setup. The figure shows a vibrating mode with two antinodes, with locations $x_{1}$  and $x_{2}$, respectively. Their respective uncertainties are denoted by $\Delta x_{1}$ and $\Delta x_{2}$. The difference $(x_{2} - x_{1})$ corresponds to $\lambda/2$.}
    \label{fig5}
\end{figure}

The complete experimental setup is shown in figure \ref{fig4}. Notice that a horizontal support had to be added because the guitar string was too short compared to the length of the original apparatus, but this detail does not affect the purpose of this lab experience.

The measurement process for half a wavelength can be seen in figure \ref{fig5}. It consists of locating two consecutive antinodes by the difference in sound as the cart is moved along the string. The position of an antinode should be where the  highest sound volume is generated. 
Nevertheless, this position has an
uncertainty associated with the perception of minor sound differences. 

To estimate this uncertainty, we can use the fact that the perturbation in the string is of sinusoidal shape. The antidone should be located where the amplitude is maximal. As shown in figure \ref{fig5}, we can estimate the position range $\Delta x_{1}$ and $\Delta x_{2}$ where the variation of the amplitude around the antinode could be imperceptible. 

For example, if we assume we can perceive a 10\% change in the amplitude around its maximum (i.e. a decrease in the sine function from unity to 0.9) the phase must change in $\sim$25º, which corresponds to a 14\% of the half wavelength in displacement. Accordingly a 14\% error in the wavelength measurement should be acceptable for this kind of setup.


\section{Results and discussion}

\begin{table}[t]
\centering
\begin{tabular}{ |c|c|c|c| }
\hline
\multicolumn{4}{|c|}{ 
Measurements 
for a string tension $T= 3.43$ N.}
 \\
\hline
Harmonic & $f$ $\left( \mbox{Hz} \right)$ & $\lambda/2$ $\left( \mbox{m} \right)$ & $v_{ms}\left(\mbox{m/s}\right)$\\
\hline
 3 & 51.00 &  0.305 & 
 31.1\\
 4 & 68.00 &  0.207 & 
 28.2\\
 5 &  85.50 &    0.150 & 
 25.7\\
 6 & 104.00 &    0.127 & 
 26.5\\
 7  & 121.50 &   0.112 & 
 27.1\\
 \hline
\end{tabular}
\caption{Measurements of $\lambda/2$ for different frequencies taken by the blind student. The string has fixed values: $\mu=4.30 \times 10^{-3}$ kg/m and $T=3.43$ N. Consequently, 
$v_{r}\sim 28.2$
m/s}
\label{table:1}
\end{table}

\begin{table}[t]
\centering
\begin{tabular}{ |c|c|c|c| }
\hline
\multicolumn{4}{|c|}{ 
Measurements 
for a string tension $T= 4.41$ N.}
 \\
\hline
Harmonic & $f$ $\left( \mbox{Hz} \right)$ & $\lambda/2$ $\left( \mbox{m} \right)$ & $v_{ms}\left(\mbox{m/s}\right)$\\
\hline
 3 & 58.50 &  0.262 & 
 30.7\\
 4 & 77.00 &  0.213 & 
 32.8\\
 5 &  96.50 &    0.177 & 
 34.2\\
 6 & 116.50 &    0.150 & 
 35.1\\
 7  & 136.50 &   0.121 & 
 33.0\\
 \hline
\end{tabular}
\caption{Measurements of $\lambda/2$ for different frequencies taken by the blind student. Fixed values: $\mu=4.30 \times 10^{-3}$ kg/m and $T=4.41$ N. Consequently, $v_{r}\sim 
32.0$ m/s}
\label{table:2}
\end{table}

Tables \ref{table:1} and \ref{table:2} show the measurements done by the blind student in a string with density density $\mu =4.30 \times 10^{-3}$ kg/m, for two different values of string tension. Table \ref{table:1} corresponds to $T= 3.43$ N and Table \ref{table:2} to $T= 4.41$ N. 

From these data the students must extract the value of the velocity of the wave in the string, by doing a linear regression of 
$f$ versus $1/\lambda$. The slope of the line corresponds to the velocity, for each Table respectively.

\begin{figure}
    \centering
    \includegraphics[width=1.0\textwidth]{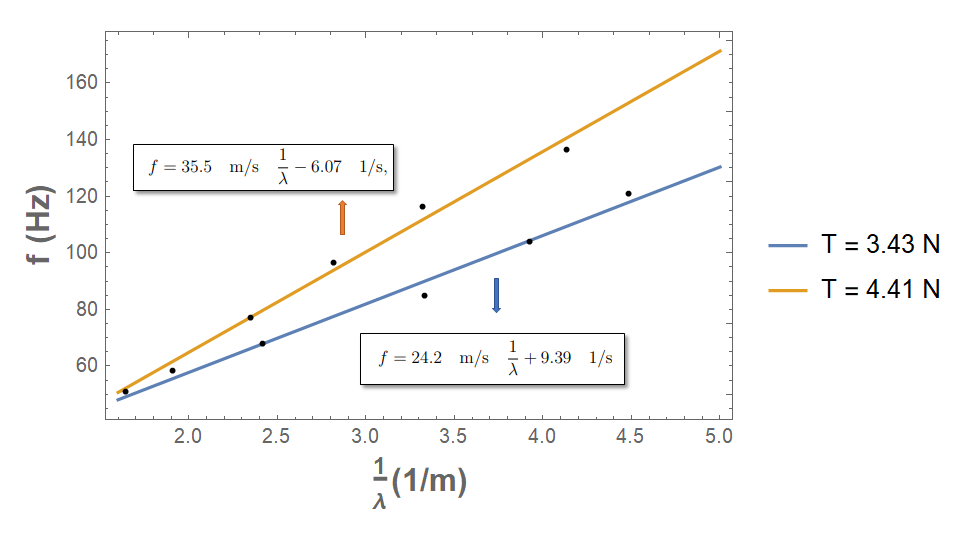}
    \caption{ Graphs of $f$ versus $1/\lambda$ of the data obtained in Table \ref{table:1} (blue dots) and Table \ref{table:2} (yellow dots). 
    The speed of the wave is obtained from the data by linear regression. 
    For the blue data, the value reported by the student was $v_{ms}= 24.2 $ m/s and for the yellow data $v_{ms}= 35.5$ m/s.}
    \label{fig6}
\end{figure}


Figure \ref{fig6} shows the data from  the two tension values mentioned above. The linear regressions reported by the student from these measurements were:

\begin{equation}
\label{firstv}
    f= 
    24.2\quad \text{m/s} \quad \frac{1}{\lambda} + 9.39 \quad \text{1/s},
\end{equation}
for the case of $T= 3.43$ N (blue dots in figure \ref{fig5}) and 
\begin{equation}
\label{secondv}
    f= 
    35.5\quad \text{m/s} \quad \frac{1}{\lambda} - 6.07 \quad \text{1/s},
\end{equation}
for the case of $T= 4.41$ N (yellow dots in figure \ref{fig5}). From Eq.~(\ref{firstv}), the student report a value $v_{ms}=24.19$ m/s and from equation (\ref{secondv})  a value $v_{ms}=35.47$ m/s. To determine the discrepancy of these velocity measurements with the theoretical value $v_r$ given in Eq.~(\ref{vref}), we calculate the fractional error defined as

\begin{equation}
    \xi=\left| \frac{v_{ms}-v_{r}}{v_{r}}\right|,
\end{equation}
where $v_{r}=
28.3$ m/s for $T=3.43$ N and $v_{r}=
32.0$ m/s for $T=4.41$ N. Consequently, the value of $\xi$ reported for $T=3.43$ N and $T=4.41$ N are $\xi=14$\% and $\xi=10$\% respectively. These error values are consistent with the experimental setup and with the error estimated in the previous section. 
We may speculate that the larger error in the first set of measurements is due to the learning process. For the second measurement, the student was more experienced, thus improving his measurement estimates.

Finally, for both measurements, an assumption when calculating $v_{ms}$ is that the measured tension does not vary during the experiment.  However, it may undergo slight variations. When the mass that generates the tension in the string is small, the string vibration induces small vertical oscillations causing the tension to suffer variations during the experiment, thus affecting the measurement process. However, we expect this error in our setup to be much smaller compared to the uncertainties in the measurement of consecutive antinodes.


\section{Conclusions}

We devised an adaptation for blind students of a teaching laboratory session dedicated to the measurement of standing waves of a string. 
The purpose of our adaptation is to allow a blind student to fully participate in the measuring process, in addition to his/her usual participation in the data analysis and report.  Our adaptation consisted in modifying the 
determination of the wavelength of the standing wave, which is usually done by visual inspection of the nodes and antinodes, 
using the sound volume generated by a guitar pickup at different points along the string. 

For the determination of the antinode position, where the most significant sound difference in the measurements occurs, a distance measurement system was built with a position sensor connected to an Arduino platform. 

The combination of these two tools
enabled a blind student to perform the half-wavelength measurements. 
The student was able to do the measurements with ease, obtaining
results within the expected error range.

\section*{Acknowledgements}

A.L. thanks Sebastian Garcia for the equipment provided and helpful discussions during the development of this work. F.J.P. acknowledges support from ANID Fondecyt, Iniciaci\'on en Investigaci\'on 2020 grant No. 11200032 and USM-DGIIE. O.N. acknowledge support from ANID PIA/Basal AFB18000. C.O.D. received support from Fondecyt (Chile) grant 1210131 and ANID PIA/APOYO AFB180002. All authors thanks to professor P. Vargas for his helpful discussions in the early stages of this study.

\vspace{1cm}



\end{document}